# Large Eddy Simulation of MILD combustion using PDF based turbulence-chemistry interaction models

ROHIT BHAYA[1], ASHOKE DE*[1], RAKESH YADAV[2]

[1]Department of Aerospace Engineering, Indian Institute of Technology Kanpur, India, 208016

[2]ANSYS Fluent India Pvt. Ltd., Pune, India, 411057

**Abstract:** The present work reports on LES of turbulent reacting flow of the Delft-Jet-in-Hot-Coflow (DJHC) burner, emulating MILD combustion, with transported PDF based combustion models using ANSYS FLUENT 13.0. Two different eddy viscosity models for LES (Dynamic Smagorinsky and Kinetic Energy Transport) along with two solution approaches for PDF transport equation, i.e. Eulerian and Lagrangian, have been used in the present study. Moreover, the effects of chemical kinetics and the micro-mixing models have also been investigated for two different fuel jet Reynolds number (Re = 4,100 and Re = 8,800). The mean velocity and turbulent kinetic energy predicted by the different models are in good agreement with experimental data. Both the composition PDF models predict an early ignition resulting in higher radial mean temperature predictions at burner exit. The models, however, correctly predict the formation mechanism of ignition kernels and the decreasing trend of the lift-off height with increasing jet Reynolds number, as observed experimentally.

*Corresponding Author: Tel.: +91-512-2597863 Fax: +91-512-2597561

E-mail address: ashoke@iitk.ac.in

*1. Introduction*

In the recent years, stringent emission norms have led to the development of new combustion technologies. The research in the area of MILD (Moderate and intense low oxygen dilution) combustion techniques have led to an efficient, low polluting combustion systems (Wünning and Wünning, 1997). According to the definition given by Cavaliere and de Joannon (2004), the inlet temperature of the reactants in case of MILD combustion process is higher than the autoignition temperature of the mixture. Further, the rise in the temperature during the combustion process is lower than the self-ignition temperature of the mixture with respect to the inlet temperature. One of the ways to achieve MILD combustion is through exhaust gas recirculation, in which the exhaust gases are mixed with the combustion air upstream before it enters into the combustion chamber. Thus, there is low oxygen concentration in the oxidizer stream, which leads to lower peak flame temperature and also helps in reducing the formation of nitric oxides. Further, with lower oxygen concentration and lower flame temperature, the chemical reactivity of the mixture reduces, making the combustion process primarily kinetics-controlled, which corresponds to low-Damköhler number regime. This type of combustion technique is also referred as HiTAC (High Temperature Air Combustion) or FLOX (Flameless Oxidation). Some of the industrial applications using MILD combustion techniques are land-base gas turbines like Alstom GT 24/26 engine (Cavaliere and de Joannon, 2004), boiler combustion chambers, domestic heating systems with flue gas recirculation and furnaces in steel industry.

Several experimental studies trying to emulate MILD combustion have been performed. Cabra et al. (2005) achieved this using a vitiated coflow burner with 12% oxygen by mass dilution in the coflow. These conditions can be considered as a limiting case for a particular burner to be classified in the MILD combustion regime. Also, laboratory-scale studies by Dally et al. (2002) and Oldenhof et al. (2010, 2011), which utilized only 3-9% and 7-11% oxygen by mass respectively in the coflow region, have been reported. The low oxygen level in the coflow led to lower temperature increase in the reaction zone and also levels of CO, OH and NO are substantially reduced. The burner which is studied experimentally by Oldenhof et al. (2010, 2011) is known as Delft-Jet-in-Hot-Coflow (DJHC) burner, and the configuration of this burner is considered to perform the numerical study in this paper.

The experimental studies by Oldenhof et al. (2010, 2011) on the DJHC burner provided a good understanding of the turbulence-chemistry interaction inside a MILD combustion regime. The burner consists of a central fuel jet of natural gas injected at a high velocity into a coflow of burnt gases produced by a secondary burner. The entrainment of hot coflow gases into the fuel jet leads to the formation of ignition kernels, due to auto-ignition of the mixture, which grow in size as they are convected downstream and thus, form a stabilized flame structure. If we increase the jet Reynolds number, the hot gases from the coflow region are entrained at a faster rate and there is a decrease in the lift-off height of the ignition kernels (Oldenhof et al., 2010). Also, increasing the coflow temperature decreases the distance between the jet exit and the location where the first ignition kernel appears. Recently, Oldenhof et al. (2013) discussed the role of turbulence on the chemical process inside the Delft flame using conditional flow field statistics and found out the existence of a sharp boundary layer, called the viscous superlayer, which separates the turbulent fuel jet region and the quasi-laminar flow region of the hot coflow gases. Across this superlayer, the OH signals drop sharply to a near zero value and that most of the chemical reactions takes places in the lean mixture fraction region just outside it. Due to the excessive dilution of the coflow, the chemical reactivity of the mixture is reduced, such that the combustion process is primarily kinetically controlled process (Damköhler number close to 1). Thus, to numerically simulate the MILD combustion regime, the chemical mechanism inside the burner should be carefully incorporated.

In recent years, several numerical studies have been performed for MILD combustion modelling. Coelho and Peters (2001) numerically studied the MILD combustion by applying an Eulerian Particle Flamelet model with RANS in an experimental burner studied by Plessing et al. (1998) and concluded that the unsteady model correctly predicted NO emissions inside the MILD combustion regime, though there were discrepancies in the results of mean and fluctuating velocity components. Christo and Dally (2005) modelled the experimental burner (Dally et al., 2002) using RANS with different combustion models – the steady flamelet model, the eddy dissipation concept (EDC) model and the mixture fraction/probability density function. Their study concluded that the EDC model predicted much better results and they stressed upon the role of differential diffusion in the MILD combustion flame. The authors extended their study by evaluating the performance of transported PDF (Christo and Dally, 2004) approach in comparison with the EDC model and concluded that the transported PDF model provided more accurate results. Kim et al. (2005) used the conditional moment closure (CMC) method to simulate the turbulent jet diffusion flame using the same burner. Their model predicted major species reasonably well but it wasn't able to account for the differential diffusion effects. The EDC model was also used by De et al. (2011) along with different two-equation RANS based turbulence model to investigate its performance for a DJHC burner (Oldenhof et al., 2010, 2011). Their model correctly predicted the experimentally observed trend of decreasing lift-off height with increasing jet Reynolds number, although, it under-predicted the lift-off height of the ignition kernels. To investigate the reason for over-prediction of mean temperature, Aminian et al. (2012) simulated the Adelaide burner (Dally et al., 2002) using RANS with EDC. They conducted the study for different turbulence model, kinetic mechanism and the effects of reactant concentration and localized extinction on the MILD flame and concluded that none of the above assertions led to higher predictions of the mean temperature and that it was a result of the combustion model. Further, De et al. (2013) performed transport PDF based modelling on the same DJHC burner setup (Oldenhof et al., 2010, 2011) and their simulations predicted better results, especially with respect to the estimations of mean temperature in comparison to their previous study (De et al., 2011). This can be attributed to the fact that the transport PDF modeling doesn't require any closure modeling for the non-linear chemical source term in its formulation.

In the context of LES modelling, Ihme et al. (2011, 2012) applied Large Eddy Simulation (LES) methodology to the Adelaide burner (Dally et al., 2002), with a steady flamelet/progress variable (FPV) model for a three-stream problem described using two mixture fractions. The first mixture fraction is associated with the mixing of the fuel and coflow and the second describes both the entrainment of surrounding air and non-uniform coflow region. They also performed calculations for three different operating conditions, in which the oxygen mass fraction in the coflow region has values 9%, 6% and 3% respectively. Their study successfully validated this three-stream burner for LES combustion models and captured the effect of different oxygen-dilution on the flame structure and suggested the use of transient scalar inflow conditions for better results. Recently, Kulkarni and Polifke (2013) simulated the DJHC flame using LES along with a combustion model based on tabulated chemistry and stochastic field modelling to take care of the turbulence-chemistry interaction. Their model satisfactorily predicted the lift-off height for different jet velocities and co-flow temperatures and captured the experimental trend of decreasing lift-off height with jet velocity but there were discrepancies in their prediction of radial velocity and temperature at different axial locations. Domingo et al. (2008) also performed LES of the vitiated coflow burner (Cabra et al., 2005) using auto-ignition and premixed-flamelet chemistry tabulation and they discussed about the significance of the formation of vortex structures inside the MILD flame near the shear layer. Also, Pope and Wang (2008), performed a Joint-PDF calculations on the Cabra H2/N2 lifted flame (Cabra et al., 2002) to study the effects of the micro-mixing models of the Lagrangian approach on the phenomena of re-ignition and auto-ignition. Their model was able to predict the formation of these events and their study also captured the lowest ignition delay time for a range of coflow temperatures. Further insight to the formation of the flame structure beginning from the mixing event was explained with the help of tracking of Lagrangian particles.

From the above review, it can be inferred that further improvement in the prediction of the statistics of the Delft flame can be achieved by accurately predicting the turbulence chemistry interaction. In order to achieve this, one of the prominent approaches utilizing the advantage of the closed chemical source term in composition-PDF based modelling can be used. Thus, LES along with PDF can prove to be a high fidelity tool to capture the ignition mechanism and development of the flame structure in the MILD flame. Therefore, the primary goal of this study is to investigate the accuracy and predictive capability of LES-PDF approach in the MILD combustion regime; in particular, to investigate the flow physics behind ignition kernel formation and flame stabilization. In this work, the experimental setup of the DJHC burner (Oldenhof et al., 2010, 2011) is considered and the simulations are carried out to demonstrate the effects of different physical parameters such as eddy viscosity models (Dynamic and KET), finite-rate chemical mechanisms (ARM9 and SKEL), the composition PDF modelling approaches (Lagrangian and Eulerian) and the effect of micro-mixing models. In addition to this, the effect of fuel jet Reynolds number on the formation of the ignition kernels and lift-off height of the Delft flames is also examined and important aspects of the formation of the flame structure are discussed. The structure of this paper is outlined as follows. The experimental setup is described first. Then information about the eddy viscosity model for LES and the different combustion models used is given. Further, description about the numerical setup is mentioned. Finally, results are compared with the experimental database and followed by the conclusion.

*2. Description of the test case*

The DJHC burner used in the current numerical study has been studied experimentally by Oldenhof et al. (2010, 2011) and is shown in Figure 1. The central fuel jet enters in a hot coflow with relatively low oxygen concentration and auto-ignition of the mixture occurs, as the temperature in the region is higher than the self-ignition temperature of the mixture. The DJHC experimental setup consists of a primary burner and a partially premixed secondary burner. The primary burner consists of a central fuel jet with D = 4.5mm, surrounded by the hot and diluted coflow gases. These coflow gases are generated upstream by the secondary burner in an annulus of diameter 82.8mm. The secondary burner is a ring burner which forms a premixed flame with additional air injected at both sides of the ring. This design of the burner yields an axi-symmetric flame and the partial premixing helps to stabilize the combustion. The central fuel jet is cooled using an air stream, which prevents excessive heating of the main fuel jet.

A series of experiments varying the coflow temperature, oxygen mass fraction and jet Reynolds number have been performed (Oldenhof et al., 2010, 2011). The numerical study conducted in this work consists of two different jet Reynolds number, as reported in Table 1 along with the boundary conditions for the fuel jet and the coflow.

## 3. Mathematical Description

The following section describes the set of governing equations used to resolve the turbulence field, the chemistry model and the turbulence-chemistry interactions.

### 3.1 Turbulence Model

In the present study, Large Eddy Simulation (LES) has been used to model the effect of turbulence in the flow field. In LES, the large structures/eddies containing the information regarding the mass, momentum, energy of the flow, are directly resolved. The segregation takes place by using a box filter, in which the flow variables are filtered in physical phase (weighted average over a given volume) (Sagaut, 2006; Pope, 2000; Poinsot, 2005). The filtered governing equations are given below–

Mass -

$$\frac{\partial \bar{\rho}}{\partial t} + \frac{\partial}{\partial x_i}(\bar{\rho}\tilde{u}_i) = 0 \qquad (1)$$

Momentum -

$$\frac{\partial \bar{\rho}\tilde{u}_i}{\partial t} + \frac{\partial}{\partial x_i}(\bar{\rho}\tilde{u}_i\tilde{u}_j) + \frac{\partial \bar{p}}{\partial x_j} = \frac{\partial}{\partial x_i}[\bar{\tau}_{ij} - \bar{\rho}(\overline{u_i u_j} - \tilde{u}_i\tilde{u}_j)] \qquad (2)$$

Chemical Species -

$$\frac{\partial(\bar{\rho}\tilde{Y}_k)}{\partial t} + \frac{\partial}{\partial x_i}(\bar{\rho}\tilde{u}_i\tilde{Y}_k) = \frac{\partial}{\partial x_i}[\overline{V_{k,i}Y_k} - \bar{\rho}(\overline{u_i Y_k} - \tilde{u}_i\tilde{Y}_k)] + \bar{\dot{\omega}}_k \qquad (3)$$

Enthalpy -

$$\frac{\partial \bar{\rho}\tilde{h}_s}{\partial t} + \frac{\partial}{\partial x_i}(\bar{\rho}\tilde{u}_i\tilde{h}_s) = \overline{\frac{Dp}{Dt}} + \frac{\partial}{\partial x_i}[\overline{\lambda\frac{\partial T}{\partial x_i}} - \bar{\rho}(\overline{u_i h_s} - \tilde{u}_i\tilde{h}_s)] + \overline{\tau_{ij}\frac{\partial u_i}{\partial x_j}} - \frac{\partial}{\partial x_i}\left(\overline{\rho\sum_{k=1}^{N}V_{k,i}Y_k h_{s,k}}\right) + \bar{\dot{\omega}}_T \qquad (4)$$

where $\overline{\frac{Dp}{Dt}} = \frac{\partial \bar{p}}{\partial t} + \overline{u_i\frac{\partial p}{\partial x_i}} \qquad (5)$

In the above governing equations, the Reynolds stresses $(u_i u_j - \tilde{u}_i \tilde{u}_j)$, species fluxes $(u_i Y_k - \tilde{u}_i \tilde{Y}_k)$, enthalpy fluxes $(u_i h_s - \tilde{u}_i \tilde{h}_s)$, filtered chemical reaction rate $\bar{\dot{\omega}}_k$, are some of the terms which need to be modelled, in order to solve the equations. The filtered laminar diffusion fluxes for species and enthalpy can be either neglected or modelled using the gradient assumption. The unresolved momentum fluxes are expressed according to the Boussineq assumption (Hinze, 1975),

$$\tau_{ij} - \frac{\delta_{ij}}{3}\tau_{kk} = -2\nu_t \overline{S}_{ij} \tag{6}$$

where $\nu_t$ is the subgrid scale turbulent viscosity and $\overline{S}_{ij}$ is the rate-of-strain tensor for the resolved case defined by –

$$\overline{S}_{ij} = \left( \frac{\partial \overline{u}_i}{\partial x_j} + \frac{\partial \overline{u}_j}{\partial x_i} \right) \tag{7}$$

The species fluxes term is modelled using a gradient assumption –

$$u_i Y_k - u_i Y_k = -\frac{\nu_t}{Sc_k} \frac{\partial Y_k}{\partial x_i} \tag{8}$$

where, $Sc_k$ is a subgrid-scale Schmidt number

In a similar manner, the enthalpy flux term is described as –

$$u_i h_s - u_i h_s = -\frac{\nu_t}{\Pr_k} \frac{\partial h_s}{\partial x_i} \tag{9}$$

where, $\Pr_k$ is the subgrid-scale Prandtl number.

In the above eqs – 6, 8, 9, the closure is obtained by modelling the unknown term $\nu_t$. These models are based on similarity assumptions, which use the information from the known large structures to estimate the effects of the smaller ones. In the present study, two different subgrid-scale eddy viscosity models: Dynamic Smagorinsky (Smagorinsky, 1963; Germano et al., 1991) and Kinetic Energy Transport (Kim and Menon, 1997), have been used to study their effect on the velocity field, which will help us to select the best model for the detailed analysis of the turbulence-chemistry interaction. The filtered chemical reaction rate term, $\bar{\dot{\omega}}_k$, is modelled using the combustion model, in this study the PDF based model, which is described in the following section.

*3.2 Turbulence-chemistry interaction model*

*Composition PDF modelling*

As stated above, PDF based modelling has been used in various studies to incorporate the chemistry inside a combustion process. The composition PDF transport equation is recast as (Pope, 1985)

$$\frac{\partial \rho f_\varphi}{\partial t} + \frac{\partial}{\partial x_i}\left[\rho u_i f_\varphi\right] + \frac{\partial}{\partial \psi_k}\left[\rho S_k f_\varphi\right] = -\frac{\partial}{\partial x_i}\left[\rho \langle u_i^{"} | \psi \rangle f_\varphi\right] + \frac{\partial}{\partial \psi_k}\left[\rho \left\langle \frac{1}{\rho}\frac{\partial J_{i,k}}{\partial x_i} | \psi \right\rangle f_\varphi\right] \quad (10)$$

where $f_\varphi$ represents the single-point, joint probability density function (PDF) of species composition and enthalpy. The first term on the left hand side represents the unsteady rate of change of PDF, the second term represents the convection in physical space by mean velocity, and the third term represents the change in $f_\varphi$ due to the chemical reactions. On the right hand side, the first term represents the transport due to velocity fluctuations and the second term represents the molecular mixing/diffusion (micro-mixing). In the above PDF transport equation, the terms on the left-hand side are closed, whereas the right-hand side terms are modelled using eddy viscosity hypothesis at the sub-grid level and micro-mixing models respectively. The main advantage of the PDF balance equation is that the chemical reaction term depends only on the chemical variables and does not require any turbulence modelling, thus, it is able to handle finite rate chemistry effects without approximations. Equation (10) can be solved either Lagrangian or Eulerian methods as described below.

*3.2.1 Lagrangian PDF method (LPDF)*

In the Lagrangian approach, the turbulent reacting flow is represented by a system of 'notional' particles whose one-point, one-time Eulerian joint PDF evolves according to the equation (10) (Haworth, 2010). The particle properties are described using stochastic differential equation. Defining the stochastic model equations corresponds to defining the closure of the PDF equation. The particles move randomly in the physical space due to the particle convection and in the composition space due to molecular mixing and reactions. The particles have mass and the sum of the masses of these particles is equal to the total mass of a control volume. The tracking of the particles is solved using fractional steps, where the states of particles are changed due to convection, mixing and reaction sub-steps.

In the convection sub-step, the particles are advanced using mean velocity from the LES field, while the micro mixing is modeled using information from mean field or the neighboring particle fields. Modeling molecular or micro-mixing term is the largest source of modeling error in PDF transport approach. In the present study, three different micro-mixing model have been used to model the molecular mixing of the species; Interaction-by-Exchange-with-Mean (IEM) (Dopazo, O'Brien, 1974), Euclidean Minimum Spanning Tree (EMST) (Subramaniam, Pope, 1998), and Coalescence Dispersion (CD) model (Janicka et al., 1978). To imitate chemistry effect in the Lagrangian approach, a GRI 2.11 based augmented reduced mechanism (ARM) kinetics, with 9 chemical species, is used. To reduce the computational cost during the time-integration of chemical reactions, the ISAT (In Situ Adaptive Tabulation) model is used (Pope, 1997).

*3.2.2 Multi-environment Eulerian PDF (MEPDF) method*

In case of MEPDF approach, the joint composition PDF transport equation is approximated using an assumed shape of the PDF, formed from a series of delta functions associated with some probability. Using this approach, the solution of equation (10) can be represented as a collection of $N_e$ Delta functions which are given as follows (Fox, 2003):

$$f_\phi(\psi;x,t) = \sum_{n=1}^{N_e} p_n(x,t) \prod_{\alpha=1}^{N_s} \delta[\psi_\alpha - <\phi_\alpha>_n (x,t)] \tag{11}$$

where $N_s$ is the number of species, $N_e$ is the number of environments, $p_n$ is the weight (or probability) of each environment, and $<\phi_\alpha>_n$ is the mean composition of any species in the $n^{th}$ environment. By substituting equation (11) into equation (10) and using the IEM micro-mixing model, the derivation described by Fox (2003), we obtain following set of governing equations which are valid for any number of environments

$$\frac{\partial \rho p_n}{\partial t} + \frac{\partial}{\partial x_i}(\rho u_i p_n) = \Gamma \nabla^2 p_n \tag{12}$$

$$\frac{\partial \rho \vec{s}_n}{\partial t} + \frac{\partial}{\partial x_i}(\rho u_i \vec{s}_n) = \Gamma \nabla^2 \vec{s}_n + C_\phi \frac{\varepsilon}{k}\left(\langle\phi_i\rangle - \langle\phi_i\rangle_n\right) + p_n S\left(\langle\vec{\phi_n}\rangle\right) + \vec{b}_n \tag{13}$$

Where, $\vec{s}_n = p_n \vec{Y}_n$ or $p_n H_n$

The transport of the probability of occurrence of $n^{th}$ environment is represented by equation (12) while equation (13) represents the transport equation of probability weighted species mass fractions or probability weighted enthalpy in each environment. The second term on the right hand side of equation (13) represents the micro-mixing term, representing the interaction between the different environments, using the IEM model and the third term is the reaction source term while the last term ($b_n$) accounts for the modeling assumptions used in the present approach and is called the correction term. The detailed derivations along with underlying assumptions are described in Fox (2003). The governing transport equations are solved for each environment using finite volume method. The reaction sources are calculated here using the same approach as in LPDF with the particle now replaced with an environment.

## 4. Numerical Setup

As stated above, the test case used in the current study is the Delft-jet-in-hot-coflow burner. The computational domain starts from the fuel jet pipe, which is situated at 10mm upstream of the jet exit, in order to form a turbulent jet flow at the pipe exit (Figure 2). Further, to consider the effects of entrainment of cold air, the domain extends till 225mm X 80mm in axial and radial directions respectively. The domain is discretized into 2.8 million grid points having grid resolution of 385 X 132 X 64 grid points in axial, radial and azimuthal directions respectively. The grid is non-uniform along axial and radial directions as shown in Figure 3. The clustering of grid points is done around the central axis in order to accurately resolve the mixing layer between fuel and the oxidizer. In the azimuthal direction, uniform meshing has been done with the generation of the O-Grid near the central jet region and can be seen more clearly in the frame zoomed at the outlet zone. The minimum and maximum filter widths are $\Delta_{min}=4.7*10^{-2}D$ (Jet exit) and $\Delta_{max}=4.1*10^{-1}D$ (Outlet). At the coflow inlet, the boundary conditions for mean velocity and its fluctuations are set equal to the experimentally observed profiles at x = 3mm (Oldenhof et al., 2010). The boundary condition for temperature is also selected using the experimental mean temperature profile measured at the same location.

However, at the fuel jet inlet, the inlet profiles (mean and fluctuations) are adjusted by a factor in order to achieve the mass conservation at jet exit location which corresponds to the experimentally measured profiles of x=3 mm. The value of species concentrations at the co-flow inlet are calculated using equilibrium assumption, which is achieved by considering the coflow to be a stream of non-adiabatic equilibrium combustion products of Dutch natural gas. The boundary condition at the outlet of the domain is set to outlet with specified pressure.

Simulations for LES are carried out using ANSYS Fluent 13.0 (2010). Two different subgrid eddy viscosity models are used: Dynamic Smagorinsky model and Dynamic Kinetic Energy Transport (KET) model. All the equations are solved with a pressure-based segregated algorithm and the convective terms in the transport equations are discretized using second order accurate bounded central differencing scheme. The pressure velocity coupling is done using the PISO algorithm. Since, the composition PDF transport approach requires closure of micro-mixing terms, the micro-mixing models used for the Lagrangian PDF approach is: IEM (Dopazo, O'Brien, 1974), EMST (Subramaniam, Pope, 1998) and CD (Janicka et al., 1978), whereas in case of Eulerian PDF, only the IEM model (Dopazo, O'Brien, 1974) is used. The chemistry is modelled using two different mechanisms, ARM9 mechanism and the SKEL mechanism (James et al., 1999). The ARM9 mechanism is reduced form of the GRI 2.11 chemical mechanism, consisting of 9 species 5 reaction steps. This reduced mechanism has shown ignition delays compared to the detailed mechanism and therefore, more resilient to flame extinction. The SKEL mechanism consists of 16 species and 41 chemical reactions steps.

## 5. *Results and Discussions*

This section first starts with the evaluation of the grid followed by the discussion on the flow field statistics. Further, the flow structure obtained in this numerical study is examined.

*5.1 Grid*

The filter width across the domain in a grid is one of the crucial aspects of LES modelling and has lot of impact on the accuracy of predictions. Therefore, the foremost task after the generation of the grid is to perform a check on its validity for the results. According to the literature, there are various mesh quality indicators to do so. According to Pope (2000), for a good LES prediction, at least 80% of the kinetic energy should be resolved. Figure 4 shows the resolution of the grid used in the current study with two different methodologies. In the first method, the ratio of turbulent kinetic energy, $k_{tur}$, and the total turbulent kinetic energy, $k_{tot} = k_{tur} + k_{sgs}$, is calculated to check for the resolution of the instantaneous fields in LES. $k_{tur}$ is calculated using the rms velocities in the domain, whereas, $k_{sgs}$ is estimated during the iteration using an external program, coupled to ANSYS Fluent through user defined functions (UDF). The UDF explicitly defines a LES test filter and then utilizing the information of grid filter size and test filter kinetic energy from the solver, it calculates $k_{sgs}$ as

$$k_{sgs} = \frac{\Delta_{sgs}}{\Delta_{test}} k_{test} \qquad (14)$$

Figure 4(a) shows the results from this formulation and it is observed that more than 85% of the kinetic energy is being resolved by the grid, specifically in the shear layer region of the domain. Further, based upon the similar principle of resolution of kinetic energy, Celik et al. (2005) proposed various indicators. One of the indicators used for this study is based on the eddy viscosity ratio:

$$LES\_IQ = \frac{1}{1 + 0.05 \left( \dfrac{\upsilon_{t,eff}}{\upsilon} \right)^{0.53}} \qquad (15)$$

In equation 15, $v_{t,eff}$ is the effective viscosity (laminar + turbulent) and $v$ is the molecular viscosity. The larger the modelled viscosity, the lower will be the quality criteria. According to Celik et al. (2005), the LES quality index should be above 0.8 for quality LES mesh. The quality of the mesh used in this study using the quality criteria of equation (15) is shown in the Figure 4(b). It can be seen, the mesh provides a good resolution of the numerical domain. Further, the energy spectrum curve obtained through FFT analysis, shown in Figure 5, is correctly able to predict the slope of -5/3 which is consistent with the literature (Pope, 2000) and suggests that the grid is able to distinguish between the integral and Kolmogorov length scales. The above tests conclude that the grid chosen for these simulations are good enough to carry out rest of the calculations.

*5.2 Flow Field Statistics*

In this section, we report the detailed discussion on flow features for varying different physical parameters.

*5.2.1 Comparison of eddy viscosity model*

After the evaluation of the grid, we now investigate the performance of the model for different eddy viscosity models. In this section, the LES with MEPDF approach using SKEL mechanism are discussed. The current result include the study of two different eddy viscosity models, i.e. Dynamic Smagorinsky (Dynamic Cs) and Kinetic Energy Transport (KET) model and two jet Reynolds numbers of Re=4100 and 8800. The numerical predictions of mean velocity ($U_x$), turbulent kinetic energy (k) and mean temperature (T) are compared with the experimental data for the different cases. Figure 6 shows the prediction for Re=4100 and it can be seen that the predictions from both eddy viscosity models are in good agreement with each other. Both the eddy viscosity models are able to correctly depict the behaviour of the flow in the shear layer (r<20mm). There is an under-prediction of the mean velocity along the centreline which is due to the over-estimation of the mean temperature. The mean temperature plots also show a peak at r = 10mm occurring at a lesser height as compared to the experimental data, which shows the sign of an early ignition occurring at that position. This is due to the over-prediction of the frequency of the ignition event, which further leads to the under-prediction in the mean lift-off height. Since both the mean velocity and turbulent kinetic energy are correctly predicted by the turbulence model, the over-estimation of the mean temperature can be attributed to the combustion model, the reasons for which will be discussed later in this study. The model predictions can be judged better by estimating the results at higher Reynolds number, which are shown in Figure 7. In case of higher Reynolds number also, the estimations from both the models are in reasonably good match to each other. Since, there is under-prediction of lift-off height for the case of 4100, increasing the jet Reynolds number would increase the effect of entrainment of the coflow gases and there would be further reduction in the lift-off height. This suggests that there is an early mixing of the jet and the co-flow gases, which in turn affects the mean axial velocity downstream. Thus, the under-estimation of the mean velocity has increased for the case of 8800 as compared to 4100. Moreover, the mean temperature plots still exhibit an onset of early ignition at X=60mm. Since no substantial differences are observed, particularly in the shear region, while using these different eddy viscosity models, the Dynamic Smagorinsky model is chosen for the rest of the simulations discussed now on unless otherwise specified, as the computational time required by it is less than the KET model (Typically Dynamic Smagorinsky takes 15% less time compared to KET model using 8 CPUs for obtaining same statistics).

As discussed earlier that due to low oxygen mass fractions, the reactions in the current burner are kinetically controlled. Therefore, it is worthwhile to check the effects of different chemical mechanism in the context of LES simulations, which is the motivation and objective of the next section.

*5.2.2 Effect of chemical mechanism*

In order to understand the effect of chemical mechanism on the accuracy of the current case, we have performed a comparative study of Re = 4100 case with MEPDF approach using two different chemical mechanism, ARM9 and SKEL. Figure 8, shows the profile of kinetic energy and axial velocity with different mechanisms. It is observed that the two mechanisms have similar predictions of axial velocity and only a minor difference in centreline turbulent kinetic energy in a short region around 40mm of axial location. The variation of mean temperature at various axial locations is shown in Figure 9. Though the results show an over-estimation of the mean temperature but there is negligible difference in the predictions by two mechanisms. The current predictions show the similar trend as reported by Aminian et al. (2012). However, a notable difference is observed during the estimation of the lift-off height of the MILD combustion. As noted in the literature (Oldenhof et al., 2010), the lift-off height is defined as the height where the formation of ignition kernels due to the auto-ignition of the mixture takes place. As a marker to define it, the first axial location of the OH mass fraction of 1e-3 is considered to be the point of an ignition event (Kulkarni and Polifke, 2013). In the present study, ARM9 predicted a lift-off height of 44mm while the SKEL mechanism estimated the lift-off height at 49mm. As stated above, the SKEL mechanism being more detailed is able to predict a better lift-off height as compared to the ARM9 mechanism. The ignition phenomena occurs at small scales in the region, i.e. region in the domain where the turbulent time scale and the chemical time scale are comparable, and thus, the present model along with the different chemical mechanism is able to capture it, although there is variation in the axial distance of its prediction with the experimental data.

It is seen in the above two sub-sections that the discrepancy between the numerical predictions and the experimental data is present with both reaction mechanisms and also with both eddy viscosity models. It can be concluded here that the combustion modelling could be another aspect which needs to be investigated to understand and explain the current discrepancies between theoretical and experimental predictions. In order to explore more on this aspect, the computations in the next section are carried out to investigate and compare the performance of different PDF transport based turbulence-chemistry interaction models for a fixed chemical mechanism.

*5.2.3 Effect of turbulence-chemistry interaction models*

Here, in this section, we discuss the predictions for the case of jet Reynolds number 4100 with ARM9 using MEPDF and LPDF solution approaches with the IEM closure. Since, LES in conjunction with SKEL mechanism is quite computationally expensive; we restrict our simulations to ARM9 mechanism in order to compare these two different PDF based models. Figure 10 shows the radial variation of turbulent kinetic energy, normal stress and shear stress and the radial variation of mean temperature and its variance for different axial locations are shown in Figure 11. It can be observed from Figure 10 that the two combustion modelling approaches led to nearly identical predictions.

In both the cases, the resolved or turbulent kinetic energy at X=15mm is under-predicted when compared to the experimental data, signifies that the effect of entrainment by the jet on the coflow gases is less in the predictions. This is more clearly visible from the Reynolds shear stress (u'v'), which show the extent of fluctuations is less. The effect of this under prediction of turbulence parameters can have an impact on the mean temperature as well. At the jet exit plane, the temperature of the fuel jet is 300K and the coflow gas maximum temperature is 1540K. As the flow convects downwards, due to the temperature gradients and entrainment effects, the mean temperature of the mixture is determined. Since at X=15mm, the effect of entrainment is less, the mean temperature predictions are in good match to the experimental data despite inaccuracies in the turbulence predictions. As we move in the downstream direction, the temperature is determined by the combined effect of entrainment and the chemistry resolution. The inaccuracies in predicting these aspects start building as we progress towards downstream as shown in the temperature profile at 60 and 90 mm locations, where we can see the over-prediction of the mean temperature. However, both the models are able to capture the profiles correctly and there are minimal differences in the two model predictions. Both the models also show a peak of the mean temperature at the downstream location of X=60mm at a radial location of 10mm, signifying the occurrence of an ignition event due to the chemical reaction between the hot coflow gases and the fuel jet. The LPDF shows slightly better match at locations away from the centreline, which is due to the fact that the MEPDF model in this region tend to approach laminar solution as supported in earlier published works (Yadav et al., 2013).

Although there is minimal difference between the two models in terms of mean temperature predictions, however, the difference between the two model predictions is prominent when we look at the fluctuations of the temperature in Figure 11 (second row), which also includes the fluctuations of the subgrid scale as well. The LPDF captures the disturbances in the shear layer more accurately than the MEPDF, which significantly under-predicted the temperature fluctuations. However, the LPDF model over-predicts the fluctuations with respect to the experimental observed values. The inaccuracy of the MEPDF in this region is due to the assumption of only 2 environments, whereas the LPDF approach involves 20 particles per cell to represent the composition PDF and hence can better predict the micro-mixing terms. The flat profile in the regions outside of the shear layer can be attributed to the scalar inflow boundary conditions used for this study. But it can also be seen that at the downstream location of 90mm, at a radial location of 20mm, the disturbances start to form again which is due to the instabilities induced in the region due to the high velocity fuel jet. Furthermore, it is noteworthy to mention that we have not considered the temperature fluctuations in the inlet BC and hence this could be one of the possible sources of errors in $T_{rms}$ predictions at upstream locations (X=15 mm) and the downstream locations as well.

Also, it is observed from Figure 11, the maximum temperature estimated by the LPDF model is 1698K which occurs at X = 90mm at a radial location of 10mm, which is 16% higher than the experimental value. For the MEPDF model, the maximum temperature is 1664K occurring at a radial location of 13mm, which is 14% higher than the experimental value. The peak temperature estimated by De et al. (2011) using the EDC model was 1795K, which on comparison with the present results suggests improved performance of the PDF models.

In is noteworthy to mention that these flames, being kinetically controlled, are more sensitive to the mixing models. Therefore, it is quite educational to investigate the effects of different micro-mixing models on the flame characteristics as discussed in the next section. Since the present formulation of MEPDF model only works with IEM micro-mixing models, the study of the effects of micro-mixing models are carried out for LPDF approach only.

*5.2.4 Effect of micro-mixing models for LPDF*

In order to understand the role of micro-mixing and the effect of change in the micro-mixing closure, the LPDF computation for Re = 4100 case are repeated in the current section with different micro-mixing models (IEM, CD and EMST) and ARM9 mechanism. In the IEM (Interaction by Exchange with the Mean) (Dopazo, O'Brien, 1974) modelling approach, the composition of all the particles in a cell are moved a small distance toward the mean composition and in the Modified Curl's / CD (Coalescence Dispersion) model (Janicka et al., 1978), a few particle pairs are selected at random from all the particles in a cell and their individual compositions are moved towards their mean composition. The EMST (Euclidean Minimum Spanning Tree) model (Subramaniam, Pope, 1998), however, takes into the effect of local mixing among the particles in the composition space and is thus, accurate in comparison to the CD and the IEM model. Figure 12, shows the radial profiles of mean and rms fluctuations of the temperature at different axial locations inside the domain. Though all the three models predict the same evolution and trend of over-estimation, a significant difference can be seen at the location of the peak temperature in the shear layer (r=10mm). The maximum mean temperature estimated by the models are observed at X = 90mm: IEM - 1698K at a radial location of 10mm, CD - 1587K at radial location of 10mm and EMST - 1588K at radial location of 12mm. The estimations show that the IEM model predicts higher region of the peak temperature at an early radial location as compared to the other two models. The peak temperature estimated by the EMST model is in well agreement with the predictions by Kulkarni and Polifke (2013). Further, the variance (second row of Figure 12) in the temperature field suggests that all the three models are able to equally predict the disturbances in the shear layer closer to the jet exit. But at an axial location of 90mm and higher, the difference in the magnitude of the predicted fluctuations is clearly visible. In the IEM model, since all the particles in the cell are moved towards their mean composition, the fluctuations in the shear layer are less and hence, the lowest predicted value of variance is recorded. The EMST model, on the other hand being local in nature, has the highest level of fluctuations among the three models. Also, the different models predicted the same lift-off height of 54mm for the MILD flame.

Having discussed the effects of eddy-viscosity models, kinetic mechanism, turbulence-chemistry interaction models and the micro-mixing models, we will venture into the details analysis of flow structures. From the above-discussed results, we consider MEPDF model along with SKEL mechanism for the detailed analysis of flow physics.

*5.3 Flow Structure*

In this section, certain features of the Delft flame like, vortex, ignition kernel, lift-off height are discussed.

*5.3.1 Formation of Vortex Structures*

Figure 13 utilizes the concept of Q-criterion method (Jeong and Hussain, 1995) to visualize the development of the turbulent jet in the flow field. The Q-criteria is defined as -

$$Q = 0.5(\Omega_{ij}\Omega_{ij} - S_{ij}S_{ij}) \tag{16}$$

where $S_{ij} = 0.5\left(\frac{\partial u_i}{\partial x_j} + \frac{\partial u_j}{\partial x_i}\right)$ and $\Omega_{ij} = 0.5\left(\frac{\partial u_i}{\partial x_j} - \frac{\partial u_j}{\partial x_i}\right)$

$S_{ij}$ refers to the magnitude of the strain rate and is the symmetrical part of the velocity gradient tensor, where-as $\Omega_{ij}$ represents the vorticity magnitude and is the anti-symmetric part of the tensor. The Q criteria represent the local equilibrium between the shear strain rate and the vorticity magnitude. In the present formulation, the result from the simulation of jet Re = 4100 using MEPDF using SKEL mechanism has been portrayed. In the, the Q Iso-surface of value $Q = 0.5\left(\frac{U_F}{D}\right)^2$ (Domingo et al., 2008), where $U_F$ is the fuel jet bulk velocity, has been shown coloured with the resolved temperature field. Due to the difference in the densities of the fuel and the coflow gases, a shear layer starts to form, leading to the onset of Kelvin-Helmholtz instability and thus, forming a vortex structure in that region. This is also noticeable from Figure 11, wherein the fluctuations of the temperature in the shear layer region ( -10mm < r < 10mm) for the LPDF model suggest that the fuel and oxidizer are mixing with each other, which eventually leads to the ignition of the mixture. Further, in Figure 13, at a downstream location of ~50mm, the formation of the Kelvin-Helmholtz toroidal structure is observed. These toroidal vortices form a ring structure and convect downstream at a reduced velocity as compared to the fuel jet velocity. It can also be seen that at various downstream location the fuel jet starts to spread out, leading to the formation of increasing toroidal vortex structures in the shear layer region and thus, it can be inferred that these vortices are a source of turbulence in the flow field. Moreover, the high temperature regions in the outer ring of the vortex structure suggest that the combustion occurs in regions of lower strain and avoids zones of intense rotation.

Figure 14 depicts, the iso-surface of spanwise vorticity coloured with the axial velocity along with streamlines coloured with temperature. At the exit of the fuel jet, the disturbance in the flow field is less. At the downstream location of ~20mm, as observed previously, the onset of breakage in the vortex structure occurs due to the Kelvin-Helmholtz instability, which keeps on increasing as we move downstream. At the same location, the low velocity hot oxidizer gases start to mix up with the high velocity fuel jet, which can also be observed from the streamline pattern. Further downstream in the domain, the vortex structure spreads out radially and starts to break down. The radial region around -10mm < r < 10mm, shows regions of low velocity which induces recirculation in the flow field and is the same region in Figure 14 where the formation of ring vortices takes place . Since this region lies inside the shear layer, the above phenomena leads to mixing between the fuel and the hot oxidizer gases, which results in the formation of a mixture of gases, having temperature higher than its self-ignition temperature. This is the fundamental principle of the MILD combustion process and leads to the formation of ignition kernels. This is also shown in the Figures 15 – 18 and occurs at the downstream location of ~30mm for this particular instant. As observed these kernels are convected downstream while growing in size and combining together to form a stabilized flame structure.

*5.3.2 Ignition Kernel and Lift-off height*

As stated by Oldenhof et al. (2010, 2011), a particular feature of the Delft flame is the formation and propagation of ignition kernels. These ignition kernels are isolated regions of flame, formed as a result of auto-ignition of the mixture due to the high temperature of the oxidizer. The kernels convect downwards, growing in size, and form a stabilized flame structure. Figures 15 and 16 represent the features of these ignition kernels obtained in this numerical study, for the two jet Reynolds number: Re = 4100 and 8800 respectively. Both the figures are made from velocity streamlines with temperature contours coloured upon them. The behaviour of the streamlines shows the effect of entrainment due to the high speed of the fuel jet and the temperature contours can give information about the ignition events. While looking closely at Figure 15, the encircled region at the time instant of t = 0.108s shows the event of formation of the ignition kernel. In the consecutive images, we see that the specific region grows in size as well as convects downwards and attaches itself to the flame front, to form a stabilized flame (t=0.119s). At the time instant of t = 0.125s, the encircled region. This shows the detailed effect of entrainment of the coflow gases by the high velocity fuel jet. This result in the generation of a shear layer across which the diffusion of the molecules takes place and leads to the development of a mixture whose temperature is higher than its self-ignition temperature. At the next instant, t = 0.126s, the occurrence of an ignition event takes place, which is independent of the earlier ignition process. The same trend is observed for the case of Re = 8800, as shown in Figure 16. Upon observing closely in Figure 16 (encircled regions), we can see the formation of recirculation regions in the shear layer occurring at an early axial location as compared to Re = 4100. This feature enhances the diffusion process and as observed from the figure this region is also convected downstream, inducing the mixing process as it moves along. The ignition time period, i.e. time between the formation of consecutive kernels, for the case of Re = 4100 varies in the range from 3ms to 12ms, while, for the case Re = 8800, the ignition time period varies from 2.4ms to 7.8ms. As stated above, due to high Reynolds number, the increasing entrainment effect leads to decrease in this ignition time period.

In addition, Figure 17 and 18 show the instantaneous OH contour at various time instant with the black line corresponding to the stoichiometric mixture fraction ($\xi_{st} = 0.07$). From the figures, it is clearly evident that the most of the reaction occurs along this particular line, especially the auto-ignition mechanism and thus, ignition kernels. To determine the lift-off height of the different cases, we would use the information obtained from the mean OH mass fraction contour (De et al., 2011; Kulkarni and Polifke, 2013). A flame pocket is defined where an OH mass fraction attains a value of 1e-3. For the case of Re = 4100, we obtain the lift-off height at 49mm and for Re = 8800, at an axial location of 41mm. It should be noted that the Figures 17 and 18 correspond to instantaneous data and we observe the mass fraction of OH attaining a value of 1e-03 even below the measured lift-off height, which is highly probable. If we compare these results from the experimental data, we observe a significant difference. This difference can be attributed to the combustion model being used. This can also be inferred from the temperature plots, that the peak in temperature at X = 60mm appears much earlier as compared to the peak of the experimental data at X = 90 mm. Thus from the above information, we can interpret that as the jet Reynolds number increases, the entrainment of the coflow gases increases, due to which the ignition time period and the lift-off height of the flame decreases. Thus, all the physical phenomena of the Delft flame are being correctly predicted by the models used in this study.

*6. Conclusions*

In this work, numerical investigation of turbulent-reacting flow of flames generated by the Delft-Jet-in-Hot-Coflow (DJHC) burner is performed using LES (Large Eddy Simulation) and Composition PDF transport model along with different conditions. First of all, the computations are performed with two different eddy viscosity models and it is found that both the models had nearly similar predictions. A similar parametric investigation with two different chemical mechanisms is done and it was observed that the two mechanisms had similar predictions of mean temperature and velocity field, but yielded different lift off height.

The SKEL mechanism having more number of species and improved reaction mechanism, compared to ARM9, is able to predict better results. It is also detected that the Lagrangian PDF is able to predict better results as compared to the Eulerian PDF as the later has more approximations. In addition, the LPDF computes the micro mixing tersm with better accuracy due to the tracking of large number particles in the domain compared to only two environments with MEPDF. In case of LPDF, it is also observed that the EMST model of the Lagrangian model shows good results in comparison to the IEM and CD models as it is able to consider the local effects in the composition space. The results are in-line with many previously reported results with pilot stabilized and bluff body flames (Cao et al., 2005, 2007).

Moreover, it is inferred from the results that the formation of vortex structure occurs due to the Kelvin-Helmholtz instability induced by the shear layer. These instabilities are a source of turbulence in the flow field which induce the formation of regions of re-circulations between the fuel and oxidizer and after an ignition delay there is an occurrence of auto-ignition, forming the ignition kernel. As observed in this study, the model used here is able to predict such phenomena and also is able to predict its convection and growth in the domain, leading to the formation of a stabilized flame structure. Moreover, upon increasing the jet Reynolds number, the model predicts a decrease in the lift-off height which is also in the agreement with the measurement.

## *7. Acknowledgements*

The support of IIT Kanpur is gratefully acknowledged as the simulations are carried out using the computers provided by them (www.iitk.ac.in/cc) and also the resources availed for data analysis and the preparation of the manuscript.

**List of Tables -**

Table 1 Boundary Conditions for the simulated cases

**List of Figures -**

Figure 1 Schematic diagram of the DJHC burner

Figure 2 Schematic diagram of the computational domain

Figure 3 Non-uniform grid of 2.8M cells used for this study. Left Frame: Magnified region of the central jet with O-grid in the domain. Right Frame: Magnified region of the domain in the shear layer

Figure 4 (a) Resolution of the turbulent kinetic energy in the shear layer of the domain and (b) LES quality criteria (Celik et al., 2005)

Figure 5 Energy spectrum showing the resolution of the inertial sub-range in LES for different Reynolds Number

Figure 6 Radial distribution of mean velocity ($U_x$), turbulent kinetic energy (k) and mean temperature (T) for the LES case of Eulerain PDF with SKEL mechanism at $Re_{jet}$ = 4100 at various downstream location. In all figures, *Symbols* are measured values and *lines* are predicted values. Δ : experimental measurements, – : SKEL-Dynamic Cs, **- -** : SKEL - KET

Figure 7 Radial distribution of mean velocity ($U_x$), turbulent kinetic energy (k) and mean temperature (T) for the LES case of Eulerain PDF with SKEL mechanism at $Re_{jet}$ = 8800 at various downstream location. In all figures, *Symbols* are measured values and *lines* are predicted values. Δ : experimental measurements, – : SKEL-Dynamic Cs, **- -** : SKEL - KET

Figure 8 Distribution of mean velocity ($U_x$), turbulent kinetic energy (k) for the LES case with Eulerian PDF with ARM9 and SKEL mechanism for $Re_{jet}$ = 4100 at various downstream

and centre line location. In all figures, *Symbols* are measured values and *lines* are predicted values. Δ : experimental measurements, – : SKEL- MEPDF, - - : ARM9 - MEPDF

Figure 9 Distribution of mean temperature (T) for the LES case with Eulerian PDF with ARM9 and SKEL mechanism for $Re_{jet}$ = 4100 at various downstream location. In all figures, *Symbols* are measured values and *lines* are predicted values. Δ : experimental measurements, – : SKEL- MEPDF, - - : ARM9 - MEPDF

Figure 10 Distribution of turbulent kinetic energy (k), Normal Stress (u'u') and Reynolds shear stress (u'v') for the LES case with ARM9 mechanism using Lagrangian PDF (LPDF) and Eulerian PDF (MEPDF) for $Re_{jet}$ = 4100 at various downstream location. In all figures, *Symbols* are measured values and *lines* are predicted values. Δ : experimental measurements, – : ARM9 - LPDF, - - : ARM9 - MEPDF

Figure 11 Distribution of the mean temperature (T) and its variance (Trms) for the LES case with ARM9 mechanism using Lagrangian PDF (LPDF) and Eulerian PDF (MEPDF) for $Re_{jet}$ = 4100 at various downstream location. In all figures, *Symbols* are measured values and *lines* are predicted values. Δ : experimental measurements, – : ARM9 - LPDF, - - : ARM9 - MEPDF

Figure 12 Distribution of the mean temperature (T) and its variance (Trms) for the LES case with ARM9 mechanism using Lagrangian PDF (LPDF) with different micro-mixing models (IEM, CD, EMST) for $Re_{jet}$ = 4100 at various downstream location. In all figures, *Symbols* are measured values and *lines* are predicted values. Δ : experimental measurements, – : ARM9 - IEM, ---- : ARM9 - CD, - - : ARM9 - EMST

Figure 13 Snapshot of the Q Iso-surface coloured with temperature (K scale) for $Q = 0.5\left(\dfrac{U_F}{D}\right)^2$, where $U_F$ is the fuel jet bulk velocity

Figure 14 Snapshot of the Iso-surface of the Axial Vorticity coloured with the axial velocity (m/s) surrounded by velocity streamlines coloured with temperature (K scale)

Figure 15 Velocity streamlines coloured with temperature contour (K scale) for the LES case with Eulerian PDF and SKEL mechanism at $Re_{jet}$ = 4100, showing the formation and convection of ignition kernels for the various time instants

Figure 16 Velocity streamlines coloured with temperature contour (K scale) for the LES case with Eulerian PDF and SKEL mechanism at $Re_{jet}$ = 8800, showing the formation and convection of ignition kernels for the various time instants

Figure 17 Instantaneous OH mass fraction distribution for $Re_{jet}$ = 4100 at various time instants. Isoline: Stoichiometric mixture fraction line ($\xi_{st} = 0.07$)

Figure 18 Instantaneous OH mass fraction distribution for $Re_{jet}$ = 8800 at various time instants. Isoline: Stoichiometric mixture fraction line ($\xi_{st} = 0.07$)

**Tables**

| Case | Jet Re | Coflow Sec. fuel (nl/min) | Coflow total air (nl/min) | $T_{max:co}$ (K) | $Y_{O2:co}$ (%) |
|---|---|---|---|---|---|
| DJHC-I_S | 4100 | 16.1 | 224 | 1540 | 7.6 |
| DJHC-I_S | 8800 | 14.2 | 239 | 1540 | 7.6 |

Table 1

**Figures**

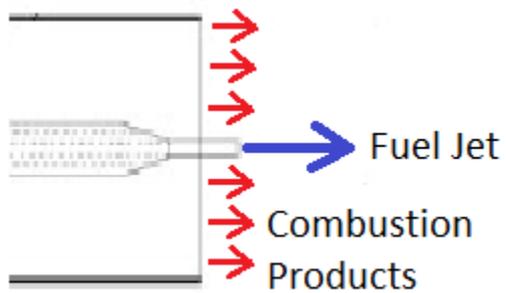

Figure 1

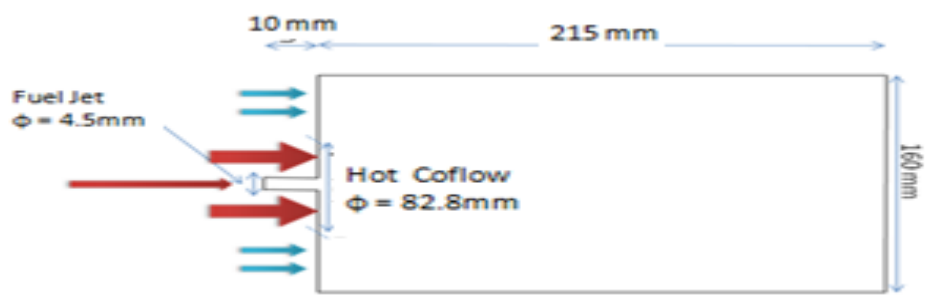

Figure 2

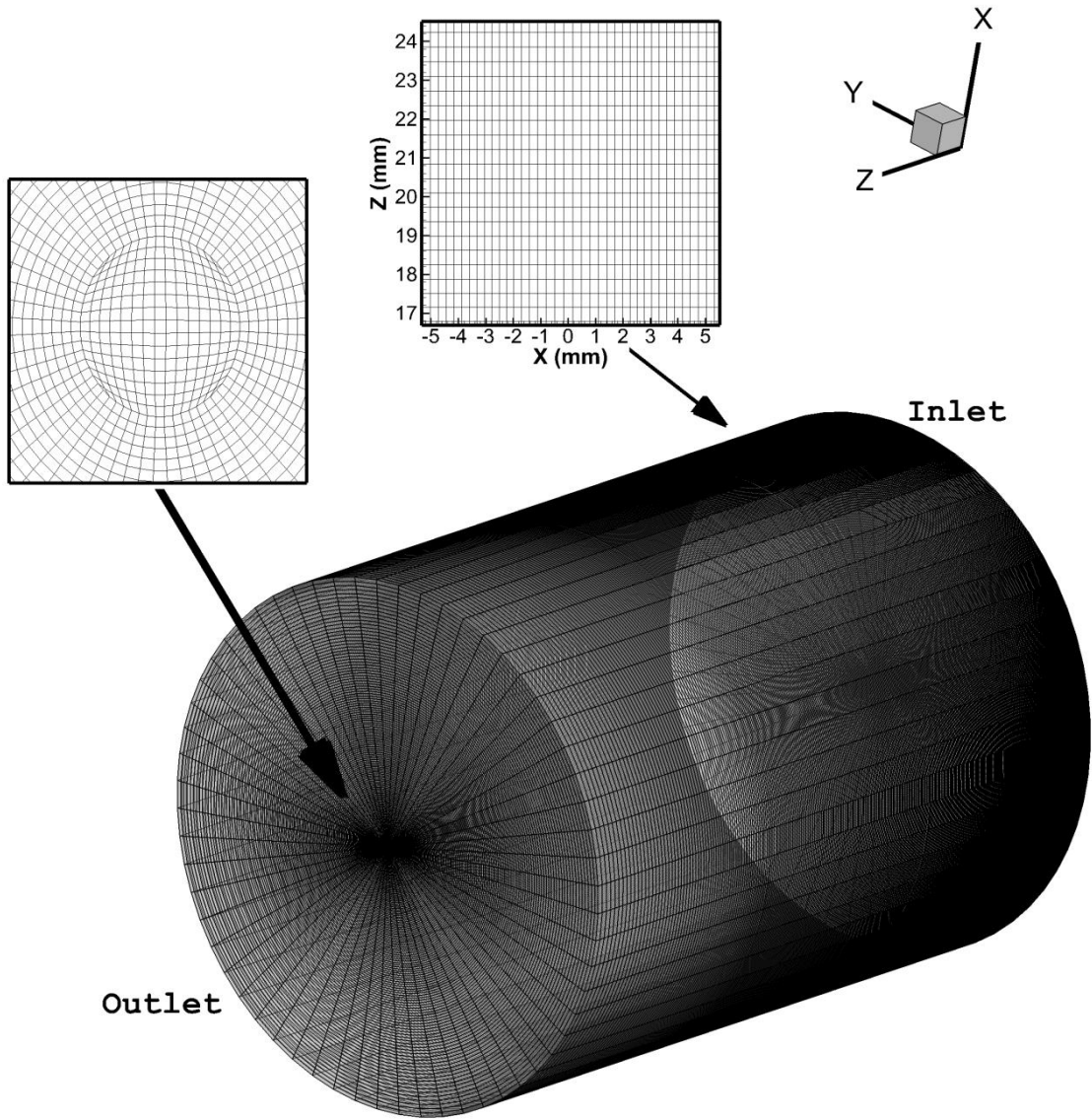

Figure 3

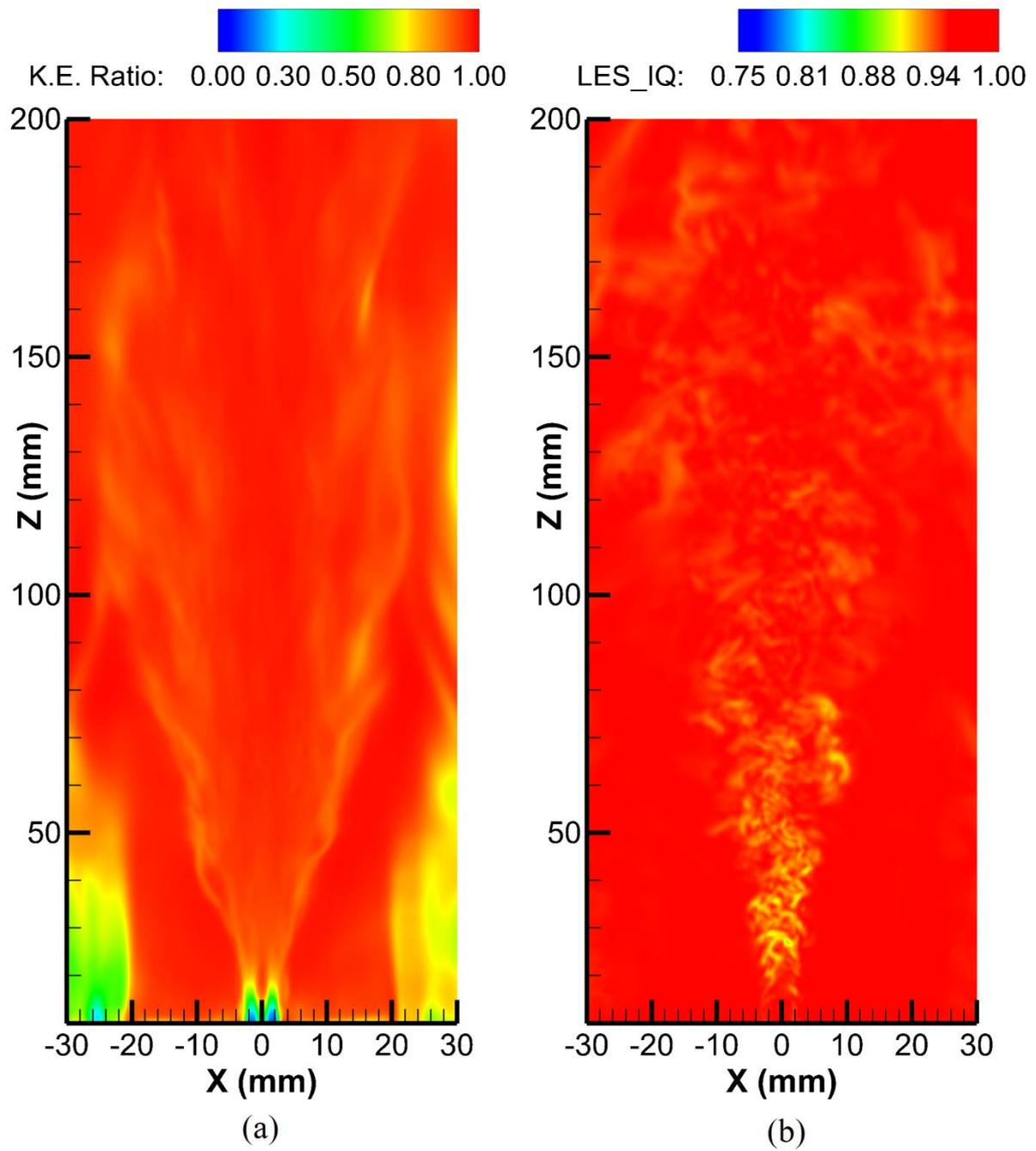

Figure 4

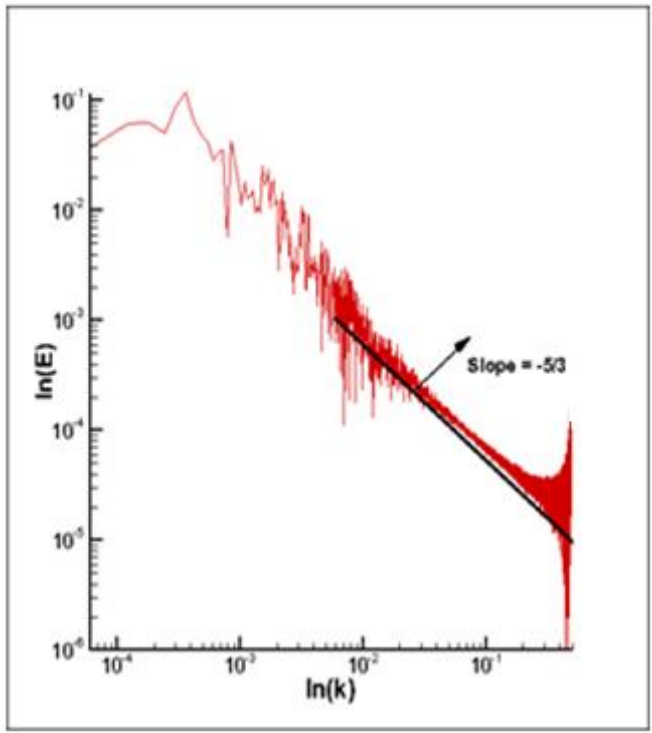

Figure 5

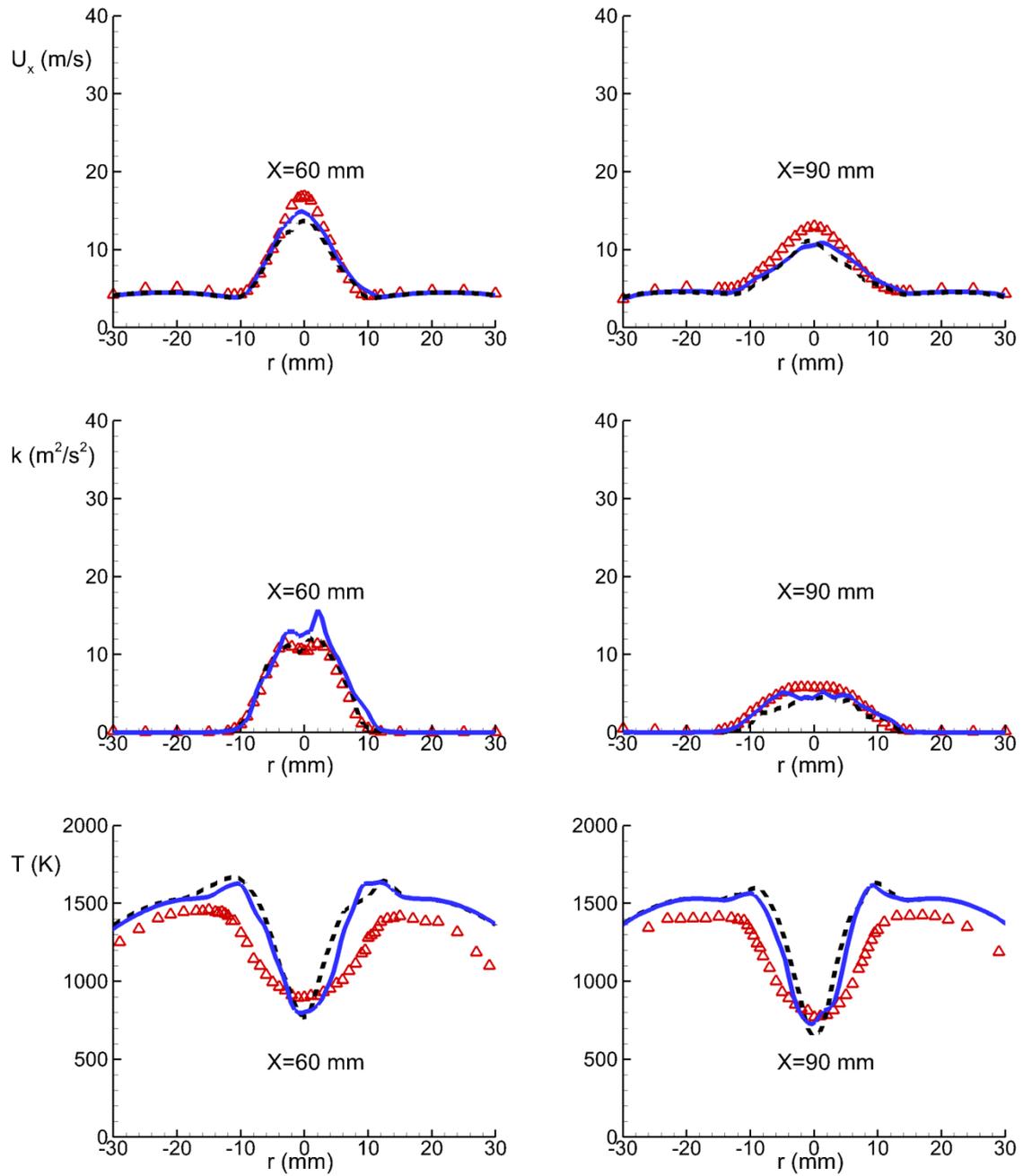

Figure 6

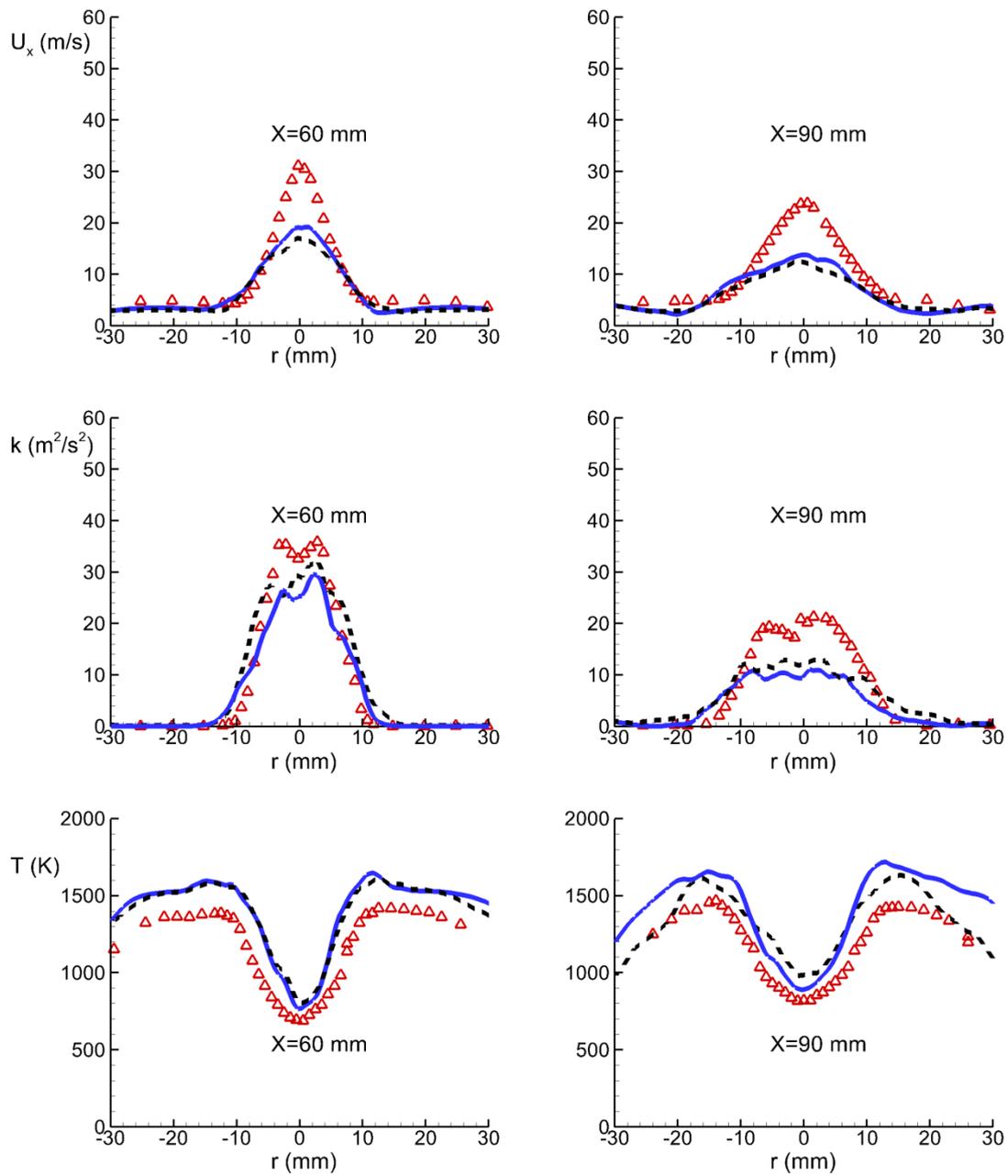

Figure 7

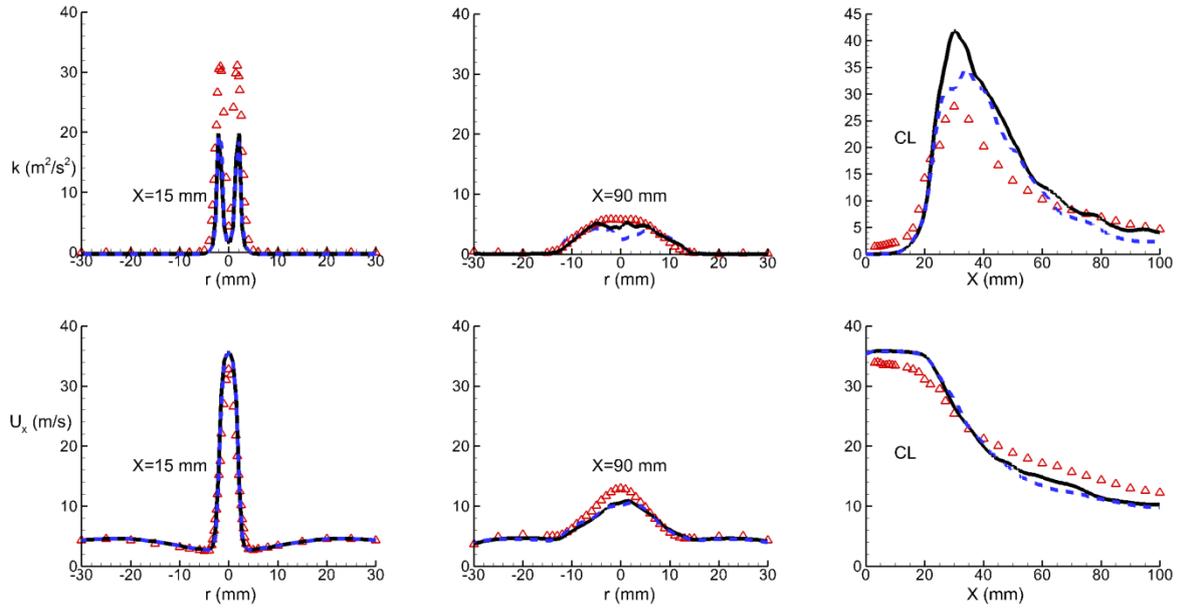

Figure 8

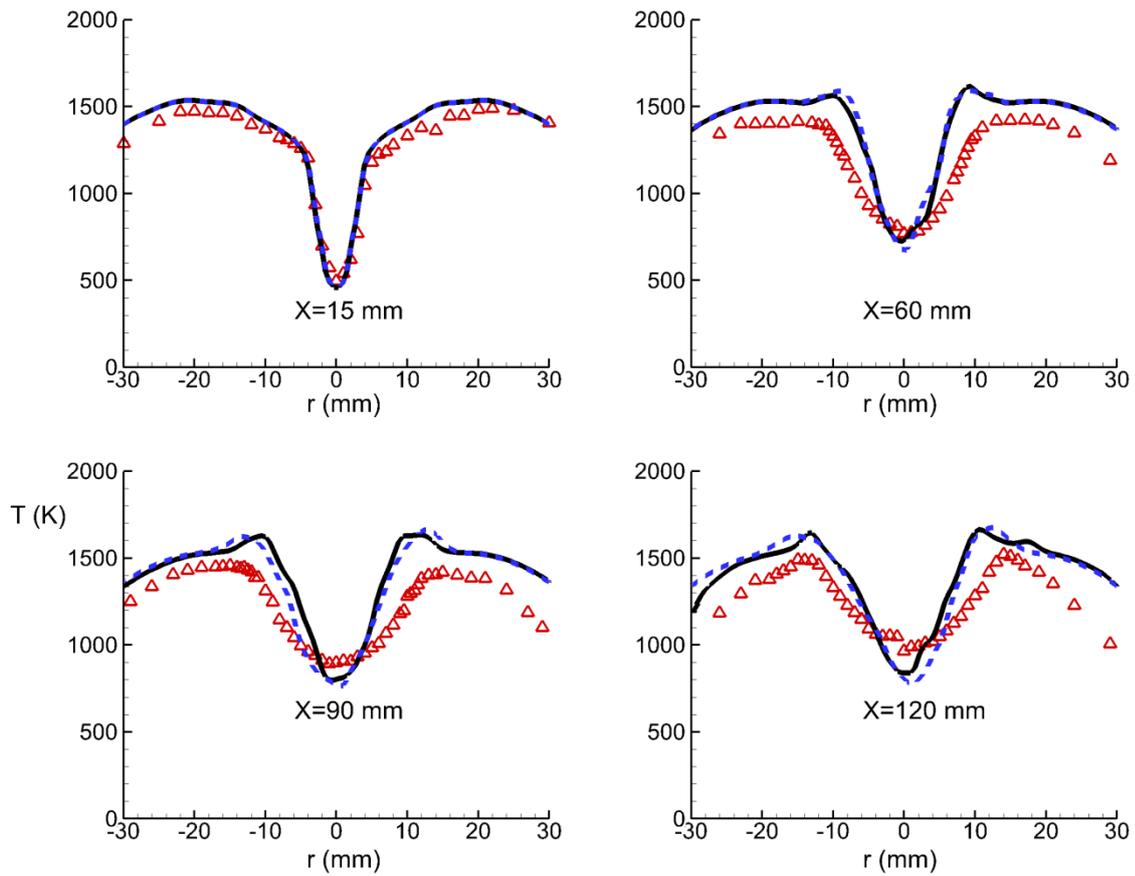

Figure 9

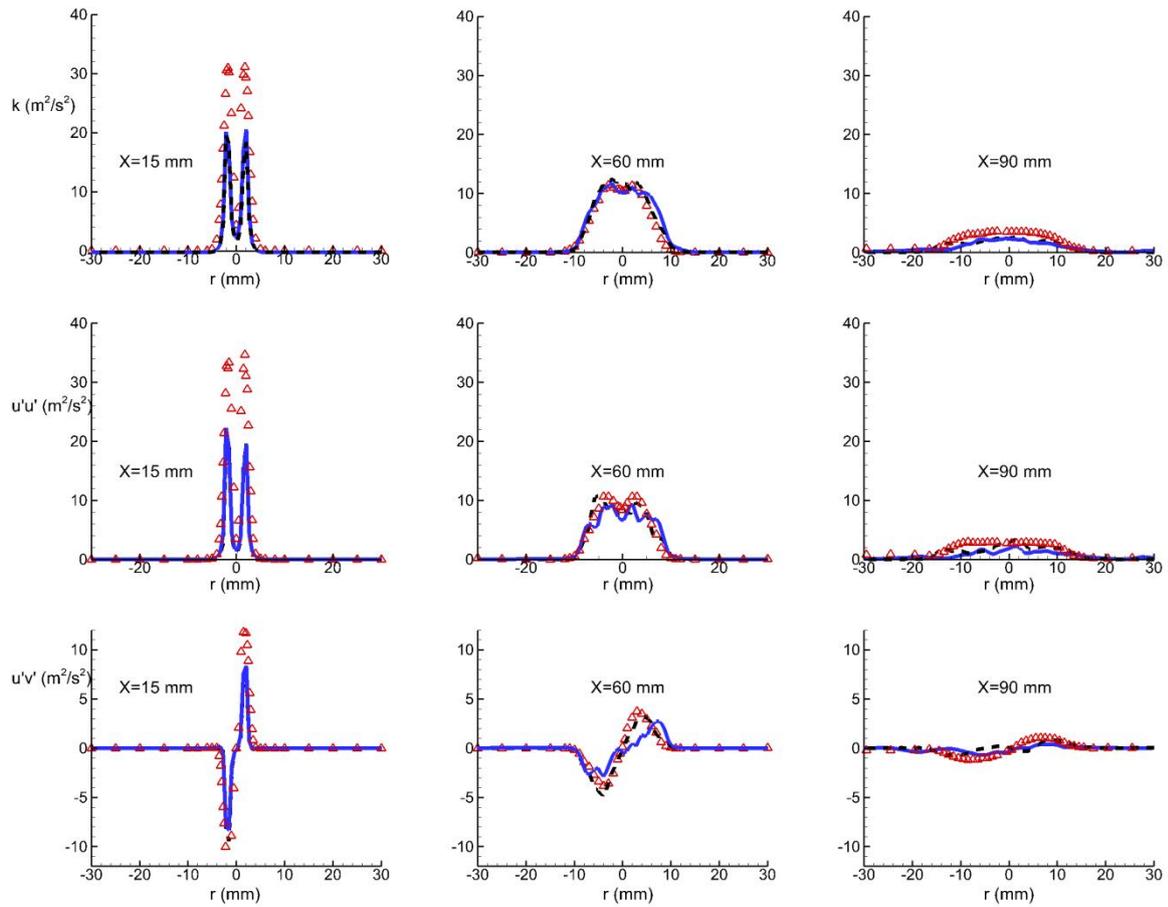

Figure 10

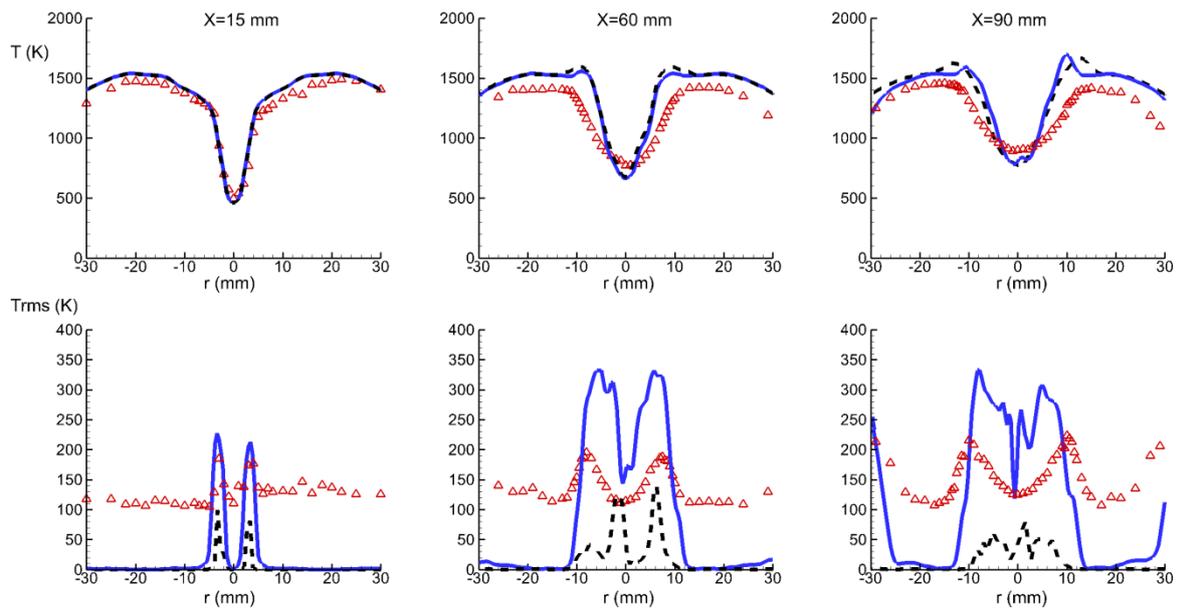

Figure 11

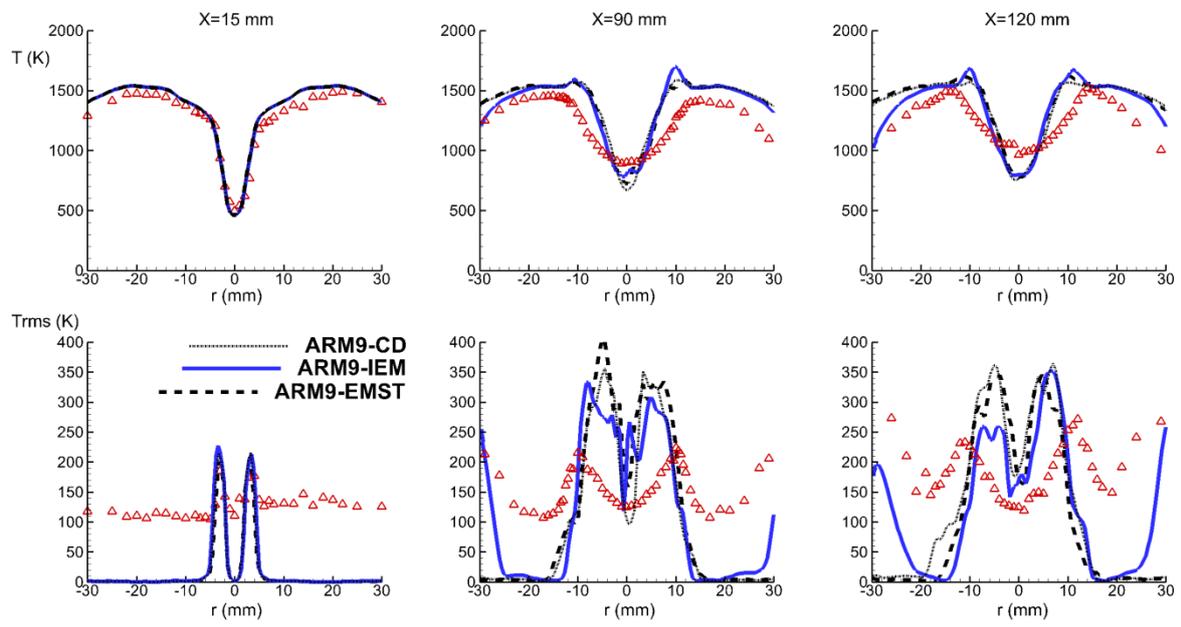

Figure 12

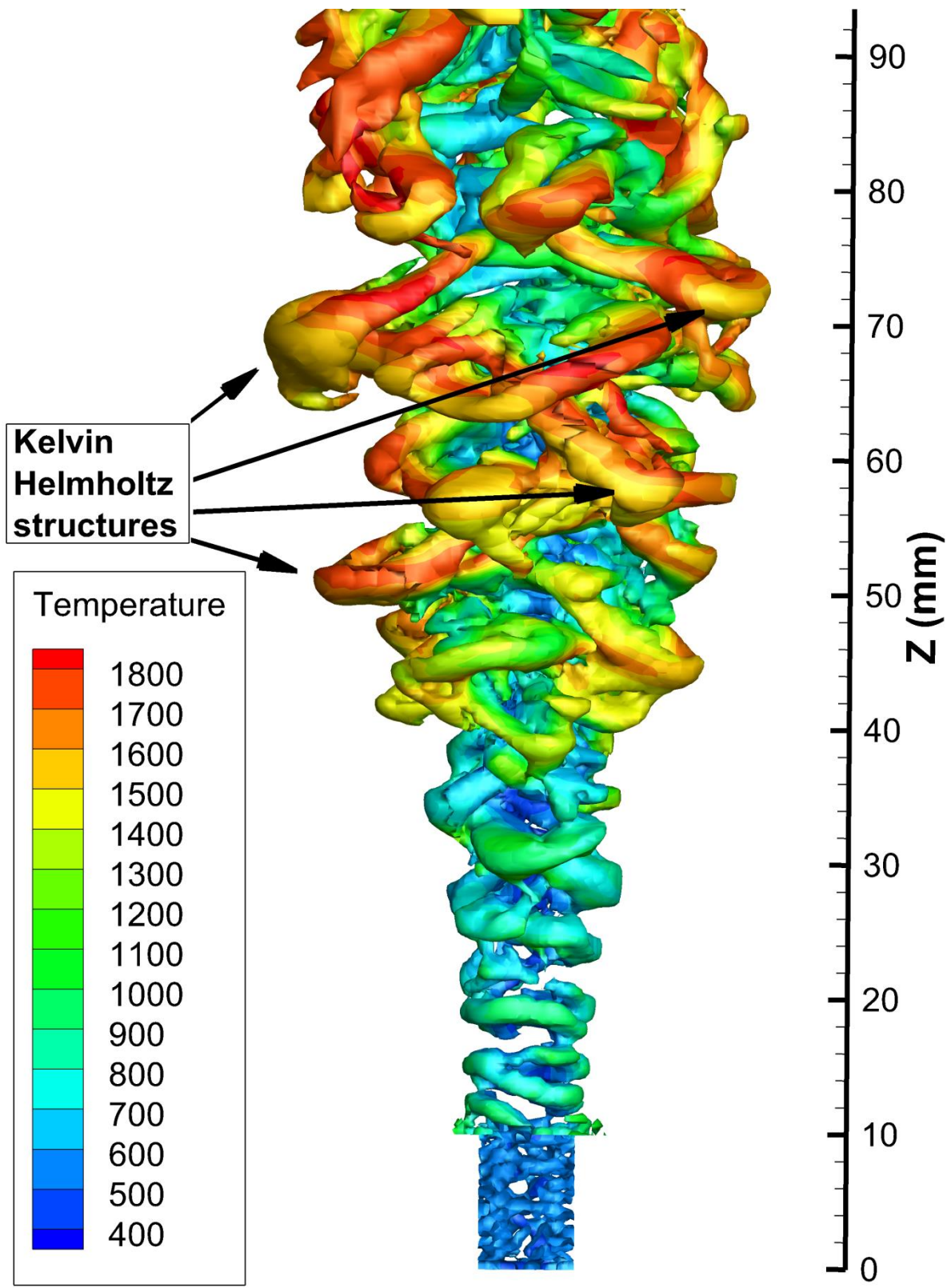

Figure 13

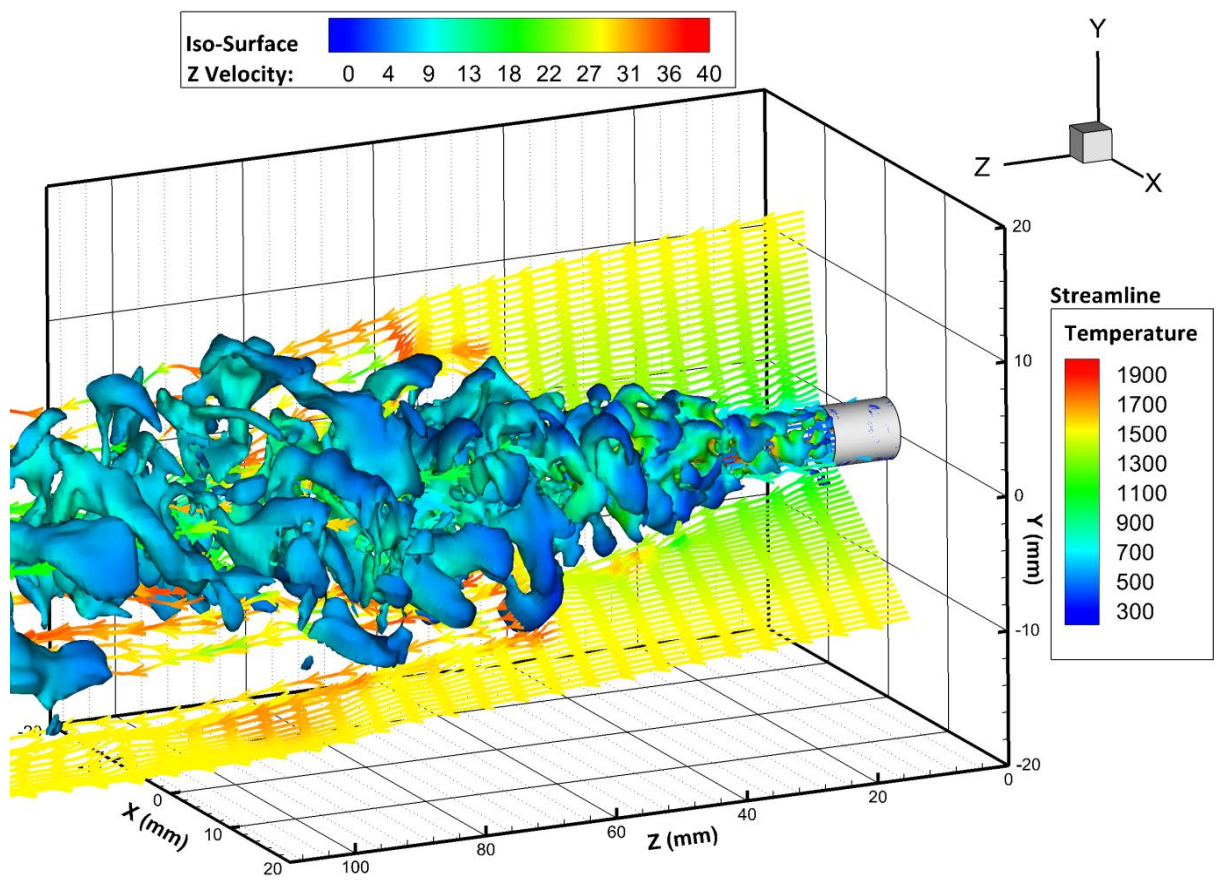

Figure 14

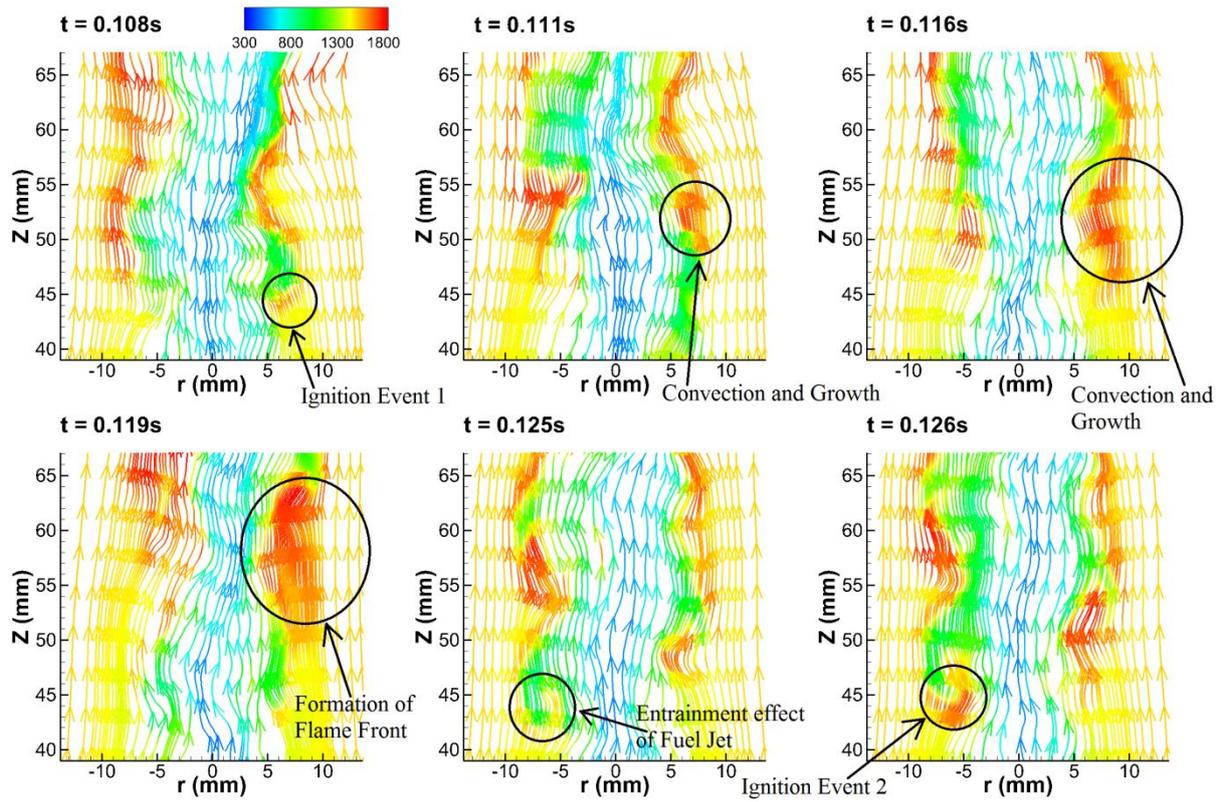

Figure 15

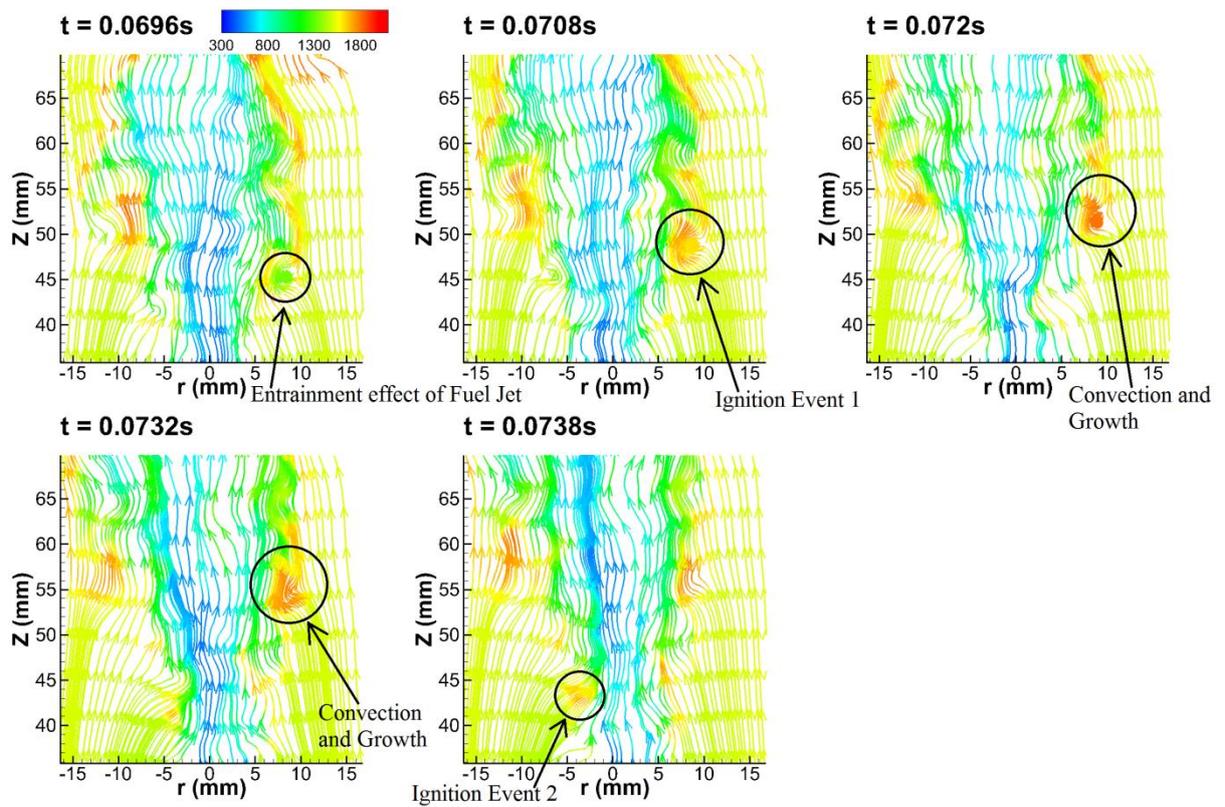

Figure 16

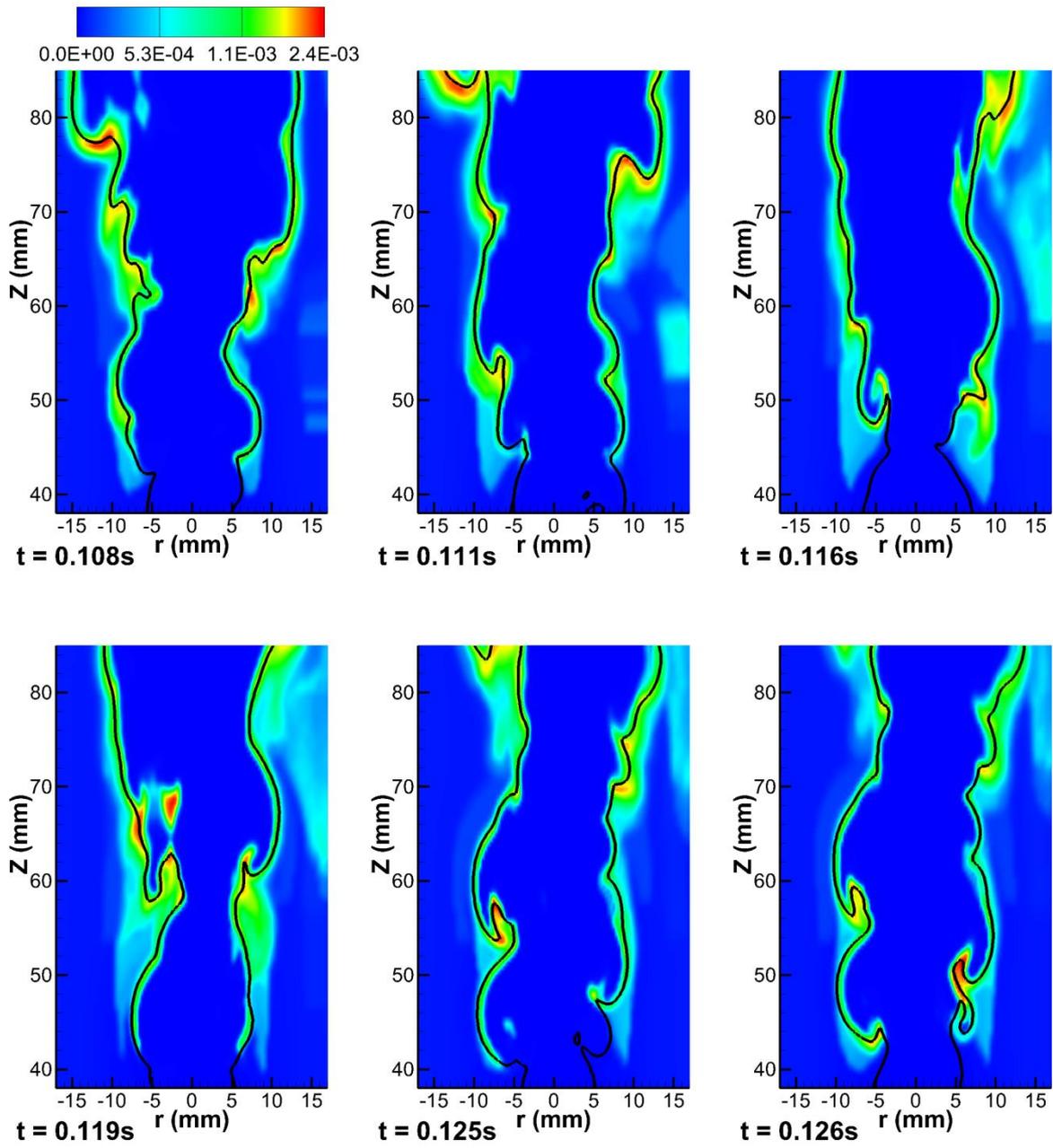

Figure 17

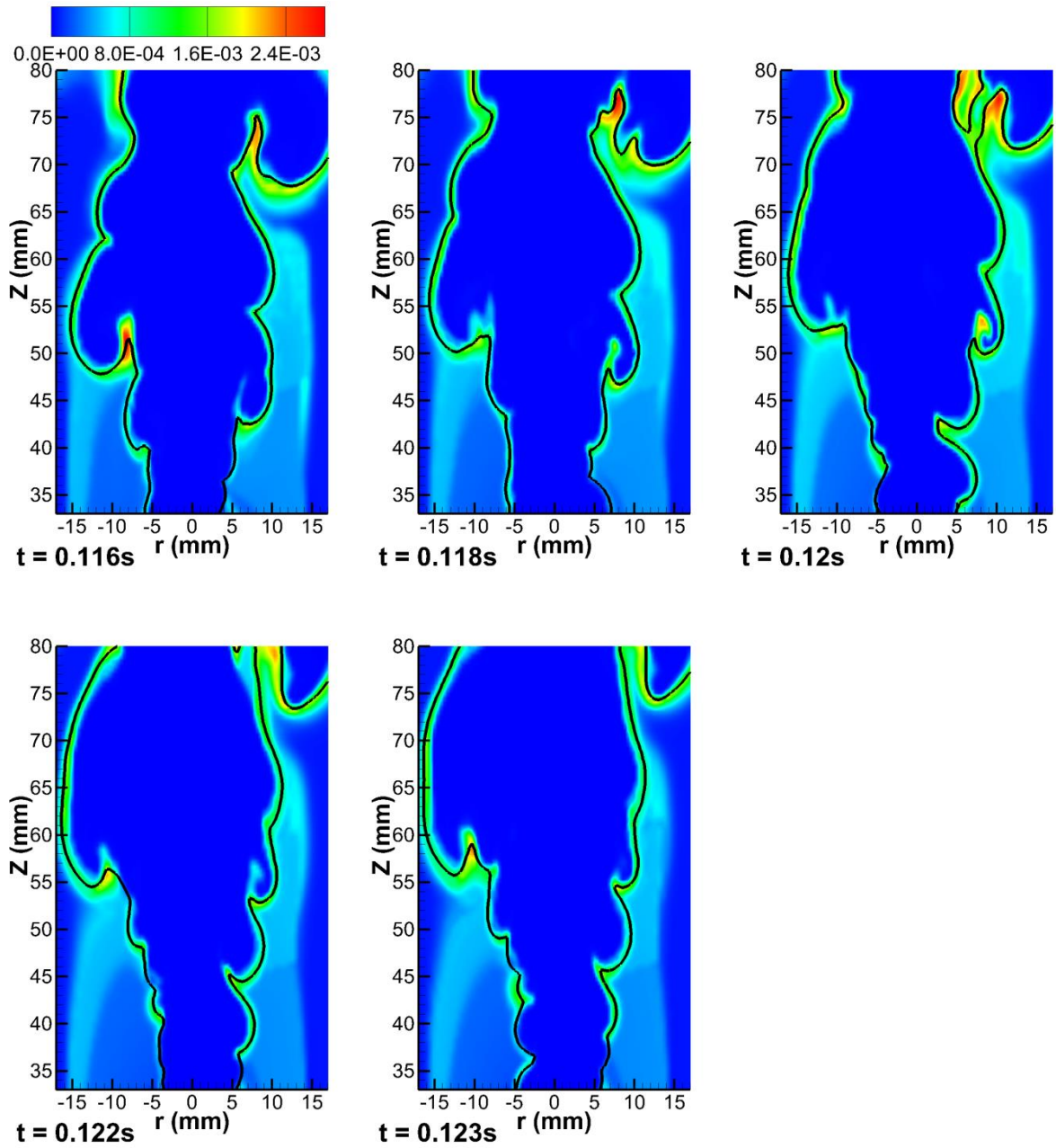

Figure 18